\documentclass[aps,pra,twocolumn,epsfig,showpacs]{revtex4}
\usepackage{dcolumn}
\usepackage{bm}
\usepackage{graphicx}
\usepackage{amsmath}
\usepackage{latexsym}
\usepackage{amsfonts}
\usepackage{amssymb}
\usepackage{array}
\usepackage{epsfig}

\newcommand{\ket}[1]{\left\vert#1\right\rangle}
\newcommand{\bra}[1]{\left\langle#1\right\vert}

\begin{document}

\title{Faithful test of non-local realism with entangled coherent states}
\author{Chang-Woo Lee$^1$, Mauro Paternostro$^2$,
and Hyunseok Jeong$^1$}

\affiliation{
$^1$Center for Macroscopic Quantum Control \& Department of Physics and Astronomy,
Seoul National University, Seoul, 151-747, Korea\\
$^2$School of Mathematics and Physics, Queen's University, Belfast BT7 1NN, United Kingdom}
\date{\today}

\begin{abstract}
We investigate the violation of Leggett's inequality for non-local realism using entangled 
coherent states and various types of local measurements. We prove mathematically the relation 
between the violation of the Clauser-Horne-Shimony-Holt form of Bell's inequality and Leggett's 
one when tested by the same resources. For Leggett inequalities, we generalize the non-local 
realistic bound to systems in Hilbert spaces larger than bidimensional ones and introduce an 
optimization technique that allows to achieve larger degrees of violation by adjusting the 
local measurement settings. Our work describes the steps that should be performed to produce a 
self-consistent generalization of Leggett's original arguments to continuous-variable states.
\end{abstract}

\maketitle

\section{Introduction}

The concepts of locality and realism at the core of Bell's celebrated inequality \cite{bell} 
and the consequences of the apparent failure of such intuitively reasonable assumptions in the 
quantum mechanical description of nature have been at the focus of a very intense theoretical 
and experimental activity \cite{BellRecent}.
Yet, it remains unclear whether the departure of a quantum mechanical entangled state from 
classicality as signaled by the violation of a Bell inequality is the result of the failure of 
locality, realism or both of them. In 2003, Leggett attempted to shed some light into this 
point by formulating an inequality that, by allowing for a degree of non-locality, tests the 
break-down of realism in an entangled resource-state \cite{Leggett}. This work has generated a 
wealth of experimental and theoretical studies directed towards the falsification of non-local 
realism with only weak assumptions on the properties of the resource state to use for the test 
and thus increasingly more experimental-friendly setups \cite{exp1,exp2,exp3}. Yet the 
investigation so far has been limited almost entirely to the case of discrete-variable states.

However, continuous variable (CV) states are endowed of interesting fundamental properties 
that, frequently, go beyond the mere extension to infinite dimensionality of those 
characterizing discrete-variable ones \cite{vLB}. Non-locality tests have been designed for 
resources belonging to this realm of quantum states \cite{Aspect,Ou,vLB,BW},
and the central role played by CV systems in photon-based quantum technology is now well 
appreciated. In particular, the class of entangled coherent states (ECSs) \cite{ECS} has 
emerged as a genuinely useful set of entangled states having a prominent role, for instance, in 
quantum
teleportation and quantum 
computing~\cite{vanEnkPRA2001,JeongPRA2001,Wang2001,JeongQIC2002,JeongPRA2002,
RalphPRA2003,BaAn03,LundPRL2008,BaAn09,Fiurasek10}. It is thus desirable to extend the formal 
apparatus designed so far for non-local realistic tests to the CV scenario. A few steps in this 
direction have been performed \cite{LeggettCV1,LeggettCV2}, although a more systematic approach 
is greatly needed. This is the main objective of the our work, which aims at stepping into a 
self-consistent formulation of non-local realistic bounds for CV states embodied by ECSs taking 
into account the inherent differences that such states have with respect to their 
discrete-variable counterparts. This is not exempt from difficulties, due to the sort of 
constraints imposed by Leggett's arguments to the local properties of the resource states to 
use and which should be reformulated in the CV case. As we show in our work, this leads to the 
necessity of re-deriving the bound for non-local realistic theories so as to introduce a (weak) 
dependence on the tested state itself. We illustrate this findings by considering various local 
operators and using the different versions of Leggett's original inequality put forward in 
Refs. \cite{exp2,exp3}. Finally, we thoroughly discuss the relation between violation of 
Bell-like inequalities and the corresponding falsification of non-local realistic theories by 
the same resource state. This nicely complements the suggestions given in Refs. 
\cite{LeggettCV1,LeggettCV2} and allows us to highlight an inherent {\it universality} of the 
behavior of the Leggett functions associated with ECSs under the formulation of the inequality 
given in \cite{exp2,exp3}.

It is important to spell out here the intrinsic significance of our work. While 
Ref.~\cite{LeggettCV2} marked an important step forward in the direction of extending Leggett's 
argument to the CV realm, the approach used there was only case-specific. The proposal put 
forward in this paper, on the other hand, provides a much more structured scenario. Starting 
from first principles, not only we unveil a previously overlooked feature of non-local 
realistic tests run with CV states (namely that the non-local realistic bound used for 
two-level systems may turn out to be rather
inappropriate when CV states are in order) but, more crucially, we design a
systematic procedure to generalize Leggett's arguments and calculate the non-local realistic 
constraints appropriate to given physical situations. Furthermore, with such a systematic 
approach, we are able to provide optimized Leggett functions which, for the case of entangled 
coherent states, require lower parameter thresholds to violate non-local realistic theories, 
therefore improving the results of Ref.~\cite{LeggettCV2}. Although our contribution here is 
the very first step towards the goal declared above, it paves the way to a full formalization 
of Leggett's inequality to infinite dimensional Hilbert spaces, an objective that we aim at 
pursuing in our current endeavors in this respect.

The remainder of this paper is organized as follows.
In Sec. \ref{review},
we briefly review the Leggett inequalities derived so far and prove the relation between the 
violation of the Clauser-Horne-Shimony-Holt (CHSH) form of Bell's inequality  and Leggett's 
one. In Sec. \ref{optimizing}, we prove that Leggett's function can be optimized over the 
measurement settings required by one of the forms of inequality introduced in Sec. \ref{review} 
so as to get a larger degree of violation. In Sec. \ref{generalizing}, we find that a 
paradoxical phenomenon arises
when testing Leggett's inequality test with an ECS and naively using the very same non-local 
realistic bound valid for two-dimensional system. We show a procedure that generalizes such 
bound to systems
other than spin-$1/2$ ones. In Sec. \ref{pseudo},
non-local realism is tested for ECS with pseudo-spin measurement operators
in terms of recently derived Leggett inequalities. Finally, in Sec. \ref{conclusions},
we summarize our findings highlighting the necessity of a more general test-tool
for non-local realism.

\setlength\arraycolsep{1pt}
\section{Review of Leggett inequalities}
\label{review}

\begin{figure}[t]
\centerline{ \bf{(a)} }
\centerline{\scalebox{0.29}{\includegraphics{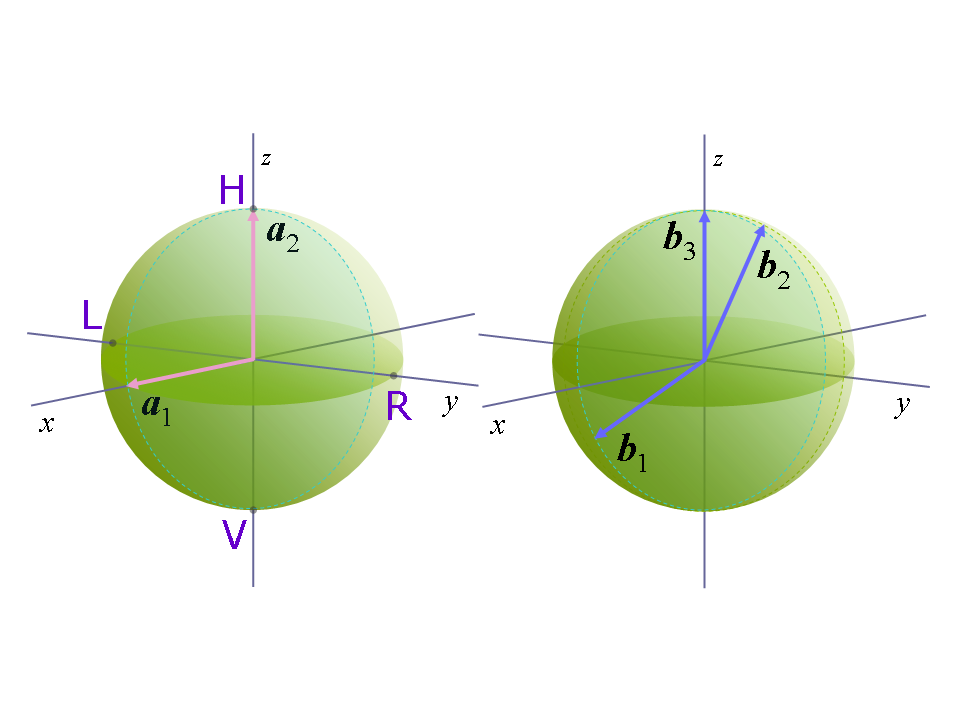}}}
\vskip-0.1cm
\centerline{ \bf{(b)} }
\centerline{\scalebox{0.29}{\includegraphics{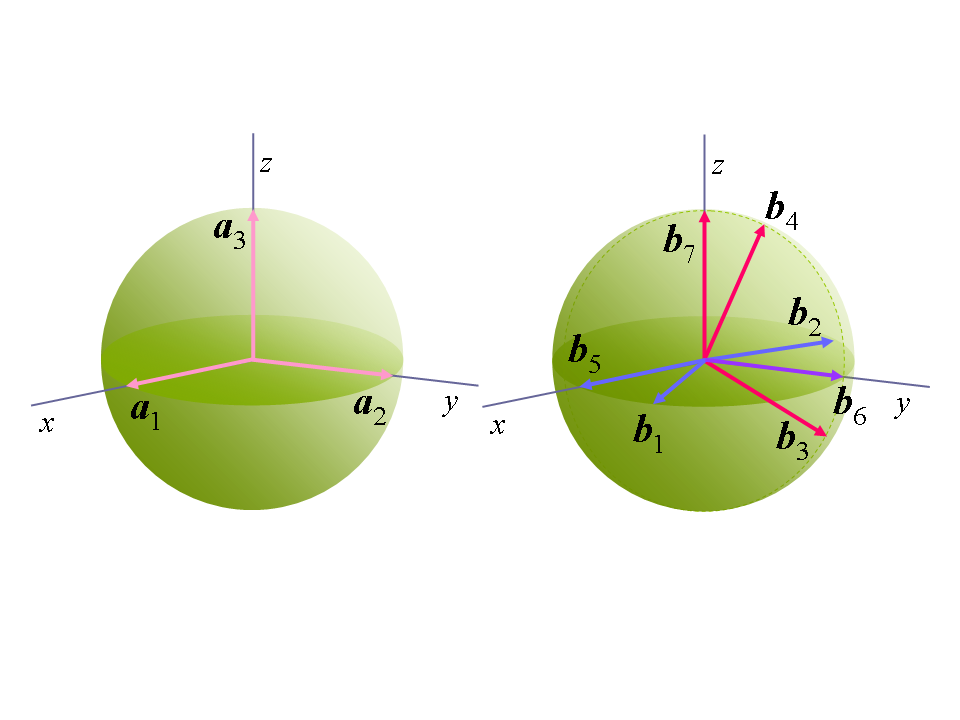}}}
\vskip-0.1cm
\centerline{ \bf{(c)} }
\centerline{\scalebox{0.29}{\includegraphics{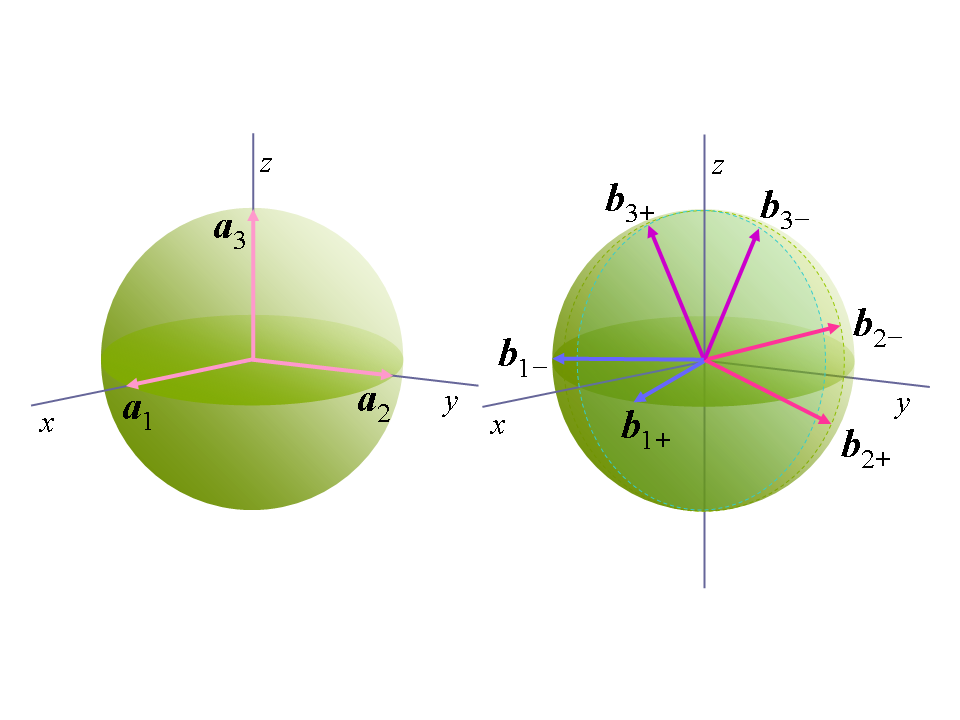}}}
\caption{(Color online)
Measurement settings for party $A$ (left) and $B$ (right)
represented on the Poincar\'{e} sphere for
{\bf (a)} the original Leggett inequality,
{\bf (b)} the``3+7 settings" Leggett inequality \cite{exp2}, and
{\bf (c)} the optimal ``3+6 settings" Leggett inequality \cite{exp3}.
In panel {\bf (a)}, H, V, R, L denote
a horizontal, vertical, right- and left-circularly
polarization respectively.
}
\label{setting}
\end{figure}

Leggett derived his inequality bearing in mind bi-dimensional systems
such as polarization states of light or spin-1/2 particles.
Although he also proposed a restricted-ensemble model
needing just a single parameter, in analogy with standard Bell's inequality tests
for the above particles, Leggett's general model exploits the full
Poincar\'{e} (for polarization states) or Bloch (for spin-1/2 states) sphere.
A restricted state-ensemble exists on a (circular) cross section
of the corresponding sphere. Referring to a local operation
as transforming a state into another on the identical sphere,
a restricted ensemble model does not necessarily require
more than one parameter for such an operation
whereas a general one requires at least two parameters. Any Leggett-type inequality starts from 
the following basic relation
\begin{equation}
\label{BasicInequality1}
\overline{A B} \leq
1 - \left| \overline{A} - \overline{B} \right|
\end{equation}
involving the arbitrary dichotomic variables $A$ and $B$ taking values $\pm 1$.
The overlines indicate statistical averages over an appropriate sub-ensemble. For instance,
in optical system this might be made out of photon states with definite polarization.
Leggett assumed that even when the subsystems are non-locally correlated,
the local average of each subsystem should fulfill {Malus' law}
\begin{equation}
\label{MalusLaw}
\begin{aligned}
\overline{A} (\mathbf{u,v;a,b}) &= \mathbf{u \cdot a} = \overline{A} (\mathbf{u;a}),\\
\overline{B} (\mathbf{u,v;a,b}) &= \mathbf{v \cdot b} = \overline{B} (\mathbf{v;b}),
\end{aligned}
\end{equation}
where $\mathbf{u}$ ($\mathbf{v}$) and $\mathbf{a}$ ($\mathbf{b}$) denote
the polarization of a photon and the measurement setting of a polarizer
at site $A$ ($B$), respectively. This relation implies that even if a non-local interaction
is allowed between subsystems $A$ and $B$, each local expectation value must depend only on
the respective local parameters.
Since $\mathbf{u}$ and $\mathbf{a}$ ($\mathbf{v}$ and $\mathbf{b}$)
can be equally represented as
(unit) vectors denoting directions of polarization on the Poincar\'{e} sphere,
a measurement vector can be transformed into another
by the same local operation as for a polarization state.
It is important to note that in Leggett's model,
the above condition imposed on local averages acts as
a constraint on the assumed non-local correlations \cite{commentosecond}.

The observation that Leggett inequality is solely based upon Eq. (\ref{BasicInequality1})
leads to prove that \emph{the violation of Leggett inequality implies violation of the CHSH 
inequality.}
We prove this by contraposition, {\it i.e.} we show that
if the CHSH inequality is not violated, Leggett inequality is not violated either.
To prove this, we start from CHSH inequality
\begin{equation}
-2 \leq \overline{A B} + \overline{A B'} +\overline{A' B} - \overline{A' B'} \leq 2,
\end{equation}
and if we set ${A'= -B'= \pm 1}$ [${A'= B'= \pm 1}$],
then we get Eq. (\ref{BasicInequality1}) [Eq. (\ref{BasicInequality2})], which proves the 
claim.

A brief construction of the original Leggett inequality is as follows.
After averaging Eq. (\ref{BasicInequality1}) over all (polarization or spin) states,
one can obtain a correlation function $E_{ij} (\varphi)$ that is bounded as
\begin{equation}\label{CorrelationFunction}
E_{ij} (\varphi) \leq 1 - f_{ij}(\varphi),
\end{equation}
where
$E_{ij}= \langle\, \overline{A_i B_j} \,\rangle$ stands for the correlation function associated 
with
measurement settings/directions $\mathbf{a}_i$, $\mathbf{b}_j$ at site $A$ and $B$ 
respectively.
While the correlation function is defined in the same way as in the standard CHSH inequality, 
the bound is not outcome independent, as  $f_{ij} = \langle\, \left| \overline{A_i} - 
\overline{B_j} \right| \,\rangle$.
This function can be considered the non-local {realistic constraint}
which, even under the presence of non-local correlations between $A$ and $B$,
limits the range of variability of the correlation function.

In order to construct the original Leggett inequality,
we choose $2$ and $3$ measurement settings for party $A$ and $B$ respectively, as illustrated 
in Fig. \ref{setting}{\bf (a)}. When written in terms of spherical polar coordinates, such 
measurement settings are given by the vectors
\begin{equation}
\begin{aligned}
&\mathbf{a}_1= \left(\frac{\pi}{2}, 0 \right), ~\mathbf{a}_2 = (0, 0),
\nonumber \\
&\mathbf{b}_1 = \left(\frac{\pi}{2} + \varphi, 0 \right), ~
\mathbf{b}_2 = \left(\varphi, \frac{\pi}{2} \right), ~
\mathbf{b}_3 = \mathbf{a}_2,
\end{aligned}
\end{equation}
where $\varphi$ is the parameter discriminating the two settings.
With these values, using Eq. (\ref{CorrelationFunction}) and taking $i=1,2$, $j=1,2,3$,
one obtains the inequality
\begin{equation}\label{inequality23}
\left| E_{11} (\varphi) + E_{23} (0)\right|
+ \left| E_{22} (\varphi) + E_{23} (0) \right|
\leq 4 - f_{\rm min} (\varphi),
\end{equation}
where $f_{\rm min}(\varphi){=}\min[f_{11}(\varphi) + f_{22}(\varphi) + 2f_{23}(0)]$
and the minimization is performed over the hidden-variable model \cite{Derivation}.
Analytically, one gets
\begin{equation}\label{constraint23}
f_{\rm min} (\varphi) = \frac{4}{\pi} \left| \sin \frac{\varphi}{2}\right|,
\end{equation}
independently of the specific hidden-variable model assumed, so that it can be applied to any 
bi-dimensional discrete system.
We have already commented on the fact that the non-local realistic bound is no longer a 
measurement-setting independent quantity, as it is very clearly exemplified by Eq. 
(\ref{constraint23}). Moreover,  due to the way the bound has been obtained, the unit vector 
defining each local subsystem in the corresponding configuration sphere remains well defined.
Gr\"{o}blacher \textit{et al.} provided the first experimental
falsification of non-local realism based on the above Leggett inequality \cite{exp1} using
the polarization-entangled state (PES)
\begin{equation}\label{PES}
\ket{\Psi^-}_{AB} =
\frac{1}{\sqrt{2}}( \ket{H}_A \ket{V}_B - \ket{V}_A \ket{H}_B)
\end{equation}
as a resource. Here $\ket{H}$ $(\ket{V})$ denotes a single-photon state with
horizontal (vertical) polarization.
They also constructed a {theoretical} non-local realistic model, based on the assumption of 
rotational invariance of the correlations arising from the use of a PES,
simulating the quantum mechanical predictions (including
 the violation of the CHSH inequality).
Paterek \textit{et al.} and Branciard \textit{et al.} \cite{exp2}
modified the original argument and elaborated new classes of inequalities that do not build on
the above-mentioned rotational invariance and require only a finite number of measurement 
settings, thus resulting in much more experimental friendly criteria \cite{exp2}.
Of those inequalities, the simplest one reads
\begin{equation}
\label{inequality37}
\begin{aligned}
L&=
\frac{1}{2} \left| E_{11}(\varphi) +E_{22}(\varphi) +E_{15}(0) +E_{26}(0) \right| \nonumber \\
&+ \frac{1}{2} \left| E_{23}(\varphi) +E_{34}(\varphi) +E_{26}(0) +E_{37}(0) \right| \nonumber 
\\
&\leq 4 - f_{\rm min}(\varphi),
\end{aligned}
\end{equation}
where $f_{\rm min}(\varphi) = \left| \sin (\varphi/2) \right|$ and
the measurement vectors shown in Fig. \ref{setting}{\bf (b)} are given by
\begin{equation}
\begin{aligned}
\label{setting37}
&\mathbf{a}_1= \left(\frac{\pi}{2}, 0 \right), ~
\mathbf{a}_2 = \left(\frac{\pi}{2}, \frac{\pi}{2} \right),~
\mathbf{a}_3 = (0,0),~\\
&\mathbf{b}_1 = \left(\frac{\pi}{2}, \varphi \right), ~
\mathbf{b}_2 = \left(\frac{\pi}{2}, \frac{\pi}{2}+\varphi \right), ~
\mathbf{b}_3 = \left(\varphi, \frac{\pi}{2} \right),~\\
&\mathbf{b}_4 = \left(\frac{\pi}{2}+\varphi, \frac{\pi}{2} \right),~
\mathbf{b}_5 = \mathbf{a}_1,~
\mathbf{b}_6 = \mathbf{a}_2,~
\mathbf{b}_7 = \mathbf{a}_3.
\end{aligned}
\end{equation}
We will refer to the above inequality as
``3+7 settings'' Leggett inequality.
It is worth noticing that, by adopting finite measurement settings
instead of infinite ones, the realistic constraint decreases and, accordingly,
the bound increases by a small amount.

In Ref. \cite{exp3}, Branciard \textit{et al.} proposed and experimentally demonstrated
the optimal inequality
\begin{equation}\label{inequality36}
L = \frac{2}{3} \sum_{i=1}^{3}
\left| E_{i+} \left(\frac{\varphi}{2}\right) + E_{i-} \left(\frac{\varphi}{2}\right)
\right|
\leq 4 - f_{\rm min}(\varphi),
\end{equation}
where $f_{\rm min}(\varphi){=}(4/3) \left| \sin (\varphi/2) \right|$
and the correlation function $E_{i\pm}$ is evaluated
for the measurement vectors $\mathbf{a}_i$ and $\mathbf{b}_{i\pm}$
shown in Fig. \ref{setting}{\bf (c)}.
Vectors $\mathbf{a}_i$'s ($i=$ 1, 2, 3) are the same as in Eq. (\ref{setting37})
while the $\mathbf{b}_{i\pm}$'s are given by
\begin{equation}
\mathbf{b}_{1\pm}{=}\left(\frac{\pi}{2}, \pm \frac{\varphi}{2}\right),
\mathbf{b}_{2\pm}{=}\left(\frac{\pi}{2} \mp \frac{\varphi}{2}, 
\frac{\pi}{2}\right),\mathbf{b}_{3\pm}{=} \left(\frac{\varphi}{2}, \frac{\pi}{2} \mp 
\frac{\pi}{2}\right).
\end{equation}
As the inequality requires
3 and 6 settings at $A$ and $B$ sites respectively,
we refer to it as the ``3+6 settings'' Leggett inequality.
The claimed optimality arises from the fact that Eq. (\ref{inequality36})  requires fewer 
settings
and has a tighter non-local realistic bound
than Eqs. (\ref{inequality23}) and (\ref{inequality37}).

\section{Generalizing Leggett-type inequalities for arbitrary systems}
\subsection{Optimizing Leggett-type inequalities by Eulerian Rotation}
\label{optimizing}

As it can be noticed by inspecting Fig. \ref{setting}, the inequalities discussed so far all 
contains
measurement settings whose vectors are parallel to the coordinate axes of the Poincar\'e 
sphere.
Such choices are due to mere algebraic convenience in the analytical derivation of the various 
forms of Leggett inequality. Moreover, for the case of PES discussed,  the correlation 
functions upon which the inequalities are built depend only on the angle between two 
measurement vectors rather than the direction of each of them individually.

\begin{figure}[b]
{\bf (a)}\hskip4.5cm{\bf (b)}
\centerline{\scalebox{0.3}{\includegraphics{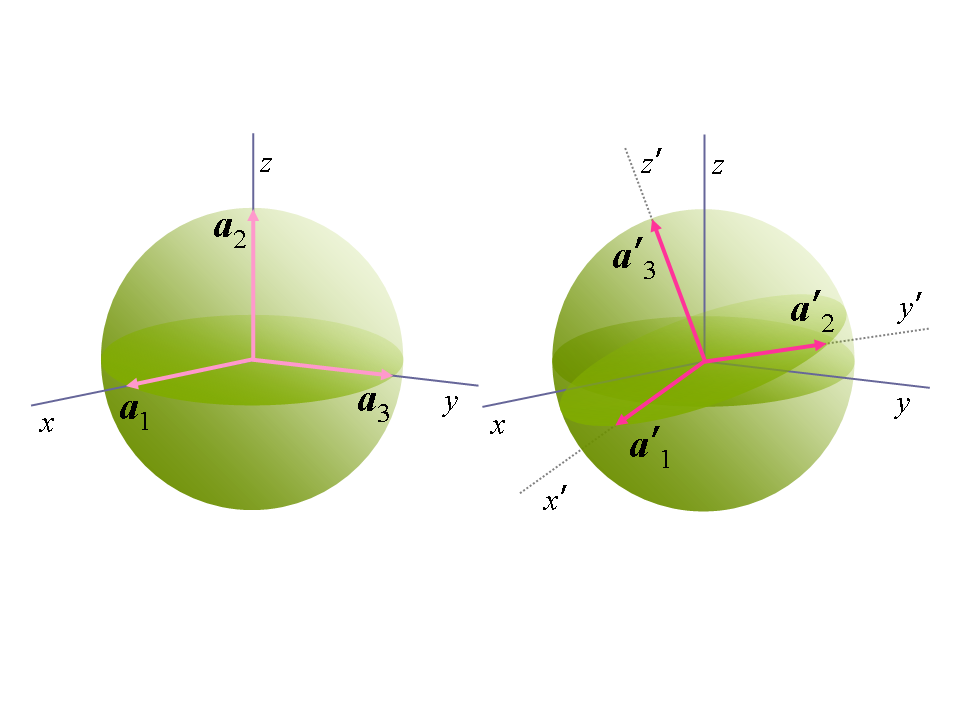}}}
\caption{(Color online)
The original measurement vectors (left) on the Poincar\'{e} (or Bloch) sphere
and their rigid-body rotated version (right). The relative angles between the
 vectors are maintained during the rotation. This approach is used in our work to optimize
 the degree of violation of non-local realistic models.}
\label{EulerRotation}
\end{figure}

Clearly, this might well not be the case for another choice of states. In what follows we show 
that, for specific resource states, by adopting a rigid-body rotation approach to the research 
for the measurement settings to be used in a Leggett-type inequality (see Fig. 
\ref{EulerRotation}),
the Leggett functions $L$'s in Eqs. (\ref{inequality23}) and (\ref{inequality37})
can be numerically optimized to get a larger degree of violation.
In contrast,  the function in Eq. (\ref{inequality36}) hardly increases under such optimization 
as
it is already optimal.

\subsection{Leggett inequality for entangled coherent states}
\label{generalizing}

So far, the approach used in both the experimental and theoretical contexts has been focused on
two-dimensional systems. However, if the state of each subsystem is well defined,
Leggett's arguments could equally well be applied to systems of larger dimensionality, such as 
qudits or continuous variables (CV). Among the states belonging to the latter class, ECSs can 
be regarded as very appealing due to their similarity to discrete states (such as the one in 
Eq. \ref{PES})
and its strong and well-studied non-local properties
\cite{BellTest1,B2,B3,BellTest2}.
In what follows we use the ECSs
\begin{equation}
\label{ECS}
\ket{\mathrm{ECS}_{\pm}}_{AB} =
\mathcal{N}_{\pm} \big[ \ket{\alpha}_A \ket{-\alpha}_B \pm \ket{-\alpha}_A \ket{\alpha}_B 
\big],
\end{equation}
where $\mathcal{N}_{\pm}$ are normalization factors.
For simplicity, we assume hereafter that $\alpha$ is real and
omit subscripts $A,~B$ denoting the two subsystems.
It should be noted that we treat the ECSs in a $2\otimes 2$ Hilbert space where the basis 
vectors are $|\alpha\rangle$ and $|-\alpha\rangle$, as in Ref.~\cite{JeongPRA2001}, even though 
they can be considered CV states.

Very recently, Leggett inequality tests on CV systems have been studied 
\cite{LeggettCV1,LeggettCV2}
adopting homodyne measurements and using the sequence of local unitary operations
\begin{equation}\label{RO}
\hat R (\theta,\phi) =
\hat D \left(-\frac{i\phi}{4\alpha} \right)
\hat U_{\rm NL} \hat D\left(\frac{i\theta}{4\alpha} \right)
\hat U_{\rm NL} \hat D\left(\frac{i\phi}{4\alpha} \right),
\end{equation}
where  $\hat a~(\hat a^\dagger)$ is the bosonic annihilation (creation) operator, $\hat 
D(\alpha) = \exp(\alpha \hat a^\dagger - \alpha^\ast \hat a)$ is the displacement operator of 
amplitude $\alpha$ and $\hat U_{\rm NL} = \exp[-i\pi (\hat a^\dagger \hat a)^2 /2]$ is the 
time-evolution operator for a field propagating in a self-Kerr medium for a dimensionless time 
$\pi/2$. The local operator $\hat R (\theta,\phi)$ transforms a coherent state $\ket{\pm 
\alpha}$ with $|\alpha|\gg{1}$ as
\begin{equation}
\label{Transformation}
\begin{aligned}
&\ket{\alpha}\rightarrow
\sin \frac{\theta}{2} \ket{\alpha} + e^{-i\phi} \cos \frac{\theta}{2} \ket{-\alpha},\\
&\ket{-\alpha}\rightarrow
e^{i\phi} \cos \frac{\theta}{2} \ket{\alpha} - \sin \frac{\theta}{2} \ket{-\alpha}.
\end{aligned}
\end{equation}
That is, $\hat R (\theta,\phi)$ mimics the effects of a rotation in the bidimensional space 
spanned by $\{\ket{\pm\alpha}\}$ and $\hat R (\theta,\phi) \ket{\alpha}$ can be any state in 
the corresponding Bloch-like sphere. Ref. \cite{LeggettCV2} used an ECS as an entangled 
resource to show that nearly the same behavior as in the discrete-system case is achieved 
\cite{LeggettCV2}.

However, CV systems may have, in general, quite different local behaviors from 
discrete-variable ones. Therefore, the same non-local realistic bound in
Eqs. (\ref{inequality23}), (\ref{inequality37}) and (\ref{inequality36}),
could not be suitable for the case of CV states too. As an example,
let us consider the general approach developed in Ref. \cite{LeggettCV2}: We retain the same 
local operations given in Eq. (\ref{RO}), but we replace the homodyne detection with an on/off 
measurement
formally described by the operator
\begin{equation}
\label{OnoffOperator}
\hat{\mathcal{O}} = \sum_{n=1}^\infty { \ket{n} \! \bra{n} }  - \ket{0} \! \bra{0}
= \openone - 2 \ket{0} \! \bra{0},
\end{equation}
where $\openone$ is the identity operator and $\ket{n}$ is a Fock state with $n$ excitations.
Such measurement has outcome $+1$ ($-1$)
if a state has \emph{any} excitations (is in the vacuum state).
The expectation value $\langle\alpha|\hat{\cal O}|\alpha\rangle$ converges to 1 as $\alpha$ 
grows
since the vacuum contribution of a coherent state diminishes accordingly.
This is the reason why the ECSs in Eq. (\ref{ECS}) show no violation of the Bell-CHSH 
inequality if $\alpha\gg{1}$ \cite{BellTest1,BellTest2}.

The correlation function for an ECS probed along the directions identified by the
 measurement vectors $\mathbf{a}_A=(\theta_A,\phi_A),~\mathbf{b}_B=(\theta_B,\phi_B)$
is given by
\begin{equation}
E_{\pm} (\varphi){=}_{AB}\!
\bra{\text{ECS}_{\pm}}
\hat\Pi_A(\theta_A,\phi_A){\otimes}\hat\Pi_B(\theta_B,\phi_B)
\ket{\text{ECS}_{\pm}}\!_{AB},
\label{eq:pat}
\end{equation}
where we have introduced the rotated on/off operators
$\hat \Pi_i (\theta_i,\phi_i) = \hat R^\dagger_i (\theta_i,\phi_i) \hat{\mathcal{O}}_i \hat 
R_i(\theta_i,\phi_i)$ ($i=A,B$). As before, $\varphi$ is the angle between $\mathbf{a}_A$ and 
$\mathbf{b}_B$. It turns out that Leggett inequality is violated for almost any value of 
$\alpha$, including the case of ${\alpha \rightarrow \infty}$, where the degree of violation 
grows to the maximum value allowed by the specific inequality being tested. This is due to the 
fact that $\langle 0 |\pm \alpha \rangle\rightarrow0$ as $\alpha\gg1$, which implies that
\begin{equation}
\begin{aligned}
\bra{\alpha} \hat\Pi_i (\theta_i,\phi_i) \ket{\alpha}
&= 1 - 2| \bra{0} \hat R (\theta_i,\phi_i) \ket{\alpha}|^2\rightarrow1,\\
\bra{\alpha} \hat\Pi_i (\theta_i,\phi_i) \ket{-\alpha}
&= e^{-2 \alpha^2} - 2 \bra{\alpha} \hat R (\theta_i,\phi_i) \ket{0}\\
&\times\bra{0} \hat R^\dag (\theta_i,\phi_i) \ket{-\alpha}\rightarrow0,
\end{aligned}
\end{equation}
regardless of the values of $\theta_i,\phi_i$.
In turn, this means that the sums in the right hand sides of
Eqs. (\ref{inequality23}), (\ref{inequality37}), and (\ref{inequality36})
all converge to $4$, thus saturating the degree of violation of the tested inequalities.

However, as explained above, the CHSH inequality cannot be violated using on/off
measurements and large-amplitude coherent states. Therefore, as proven in the previous section,
Leggett inequality cannot be violated either in the same range of $\alpha$, in striking 
contrast with what has been observed above. This paradoxical situation arises exactly from the 
reasons highlighted before, {\it i.e.} a state living in a Hilbert space that is not 
bidimensional may deviate from the predictions of  Malus' law and, as such, could originate a 
new non-local realistic constraint $f_{\rm min}(\varphi)$.
With this in mind, we have to look for a quantum-mechanical substitute of the classical Eqs. 
(\ref{MalusLaw}). As  Malus' law deals with the statistical average of the expectation value of 
an (arbitrary) measurement vector
with an (arbitrary) state on the Bloch sphere, a natural yet rigorous way to re-formulate it in 
this context is to consider the local average of the expectation value of a rotated on/off 
measurement
with a rotated coherent state. That is
\begin{equation}
\label{MalusLawOnoff}
\overline{A} (\mathbf{u;a}) =
\bra{\alpha} \hat R^\dagger (\theta_u,\phi_u) \hat \Pi(\theta_a,\phi_a)
\hat R(\theta_u,\phi_u) \ket{\alpha}
\end{equation}
with ${\bf u}=(\theta_u,\phi_u)$ and ${\bf a}=(\theta_a,\phi_a)$.
With this at hand, we could in principle newly evaluate
$f_{\rm min}(\varphi)$ for the on/off-measurement approach.
However, as it is difficult to derive a closed-form analytical expression,
we have numerically evaluated the non-local realistic bound
for each value of $\varphi$. Unfortunately,
no violations of Leggett inequalities can be observed, despite the use of the
optimization technique described in the previous Section. For on/off measurements,
the constraints turn out to be too weak to falsify non-local realistic models tested with ECSs:
The Leggett function and the bound have values nearly approaching 4 in almost all range of 
$\alpha$ and $\varphi$ and the former has slightly smaller values than the latter.

\begin{figure}[t]
\centerline{\scalebox{0.36}{\includegraphics{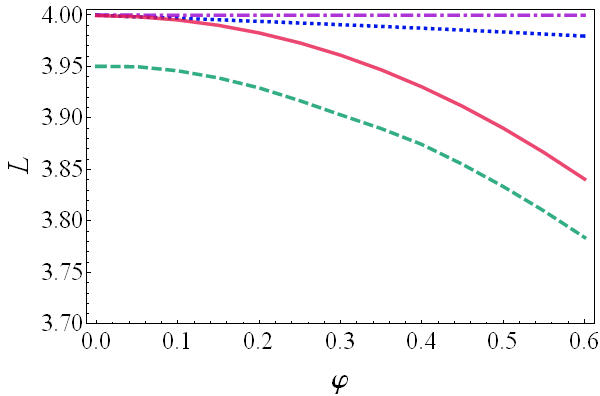}}}
\caption{(Color online)
Optimized 3+7 setting Leggett function $L$ for $\ket{\mathrm{ECS}_-}$ with parity measurement 
against $\varphi$
at $\alpha=$ 5 (green dashed) and 50 (red solid).
The non-local realistic bounds at $\alpha=$ 5 (blue dotted) and 50 (violet dot-dashed) are also 
plotted.
The plotting range is the same as Fig. \ref{LeggettTest37} for comparison.
Note that the optimized Leggett function turns out not to overcome the bound in any range of 
$\alpha$ and $\varphi$ including the above range.
}
\label{LeggettTest37Parity}
\end{figure}

To provide an example of measurement which cannot falsify the Leggett inequality,
we use the following parity operator instead
\begin{equation}\label{ParityOperator}
\hat{\mathcal{O}} = \sum_{n=0}^\infty \big[ \ket{2n+1} \! \bra{2n+1}   - \ket{2n} \! \bra{2n} 
\big],
\end{equation}
which gives $+1$ ($-1$) when a state has an odd (even) number of excitations.
The CHSH inequality can be tested via phase-space methods and using displaced parity operators 
and the Wigner function~\cite{BW,BellTest1,BellTest2}. Similarly,
one can perform Leggett inequality tests in the same way as in the previous Sections
by replacing the on/off operator in Eq. (\ref{OnoffOperator}) with the above parity operator.
However, this would not be sufficient to observe any violation.
As partly can be seen in Fig. \ref{LeggettTest37Parity},
the Leggett function cannot overcome the bound in any range of $\alpha$ and $\varphi$
even when adopting the optimization scheme in Sec. \ref{optimizing}.

\begin{figure}[b]
{\scalebox{0.36}{\includegraphics{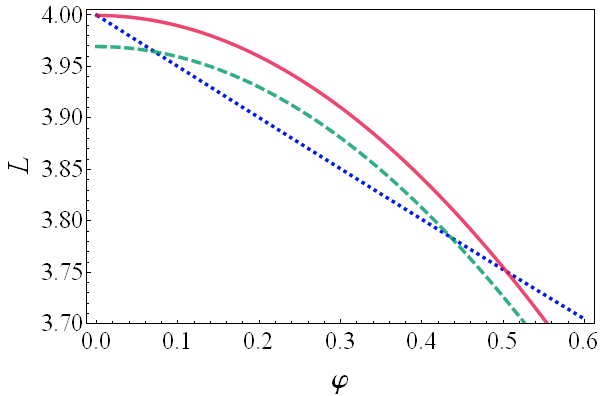}}}
\caption{(Color online)
Leggett function $L$ for $\ket{\mathrm{ECS}_-}$ against $\varphi$
at $\alpha=$ 5 (green dashed) and 50 (red solid).
The non-local realistic bound (blue dotted) is also plotted.
Note that the bounds corresponding to $\alpha=$ 5 and 50 are practically coincident. This can 
also be seen in Fig. \ref{LeggettTest37Opt} {\bf (c)} and {\bf (d)}.
The violations are maximized at $\varphi \simeq 0.25$ regardless of the actual value of 
$\alpha$.
}
\label{LeggettTest37}
\end{figure}

\begin{figure*}[t]
\hskip0.9cm{\bf{(a)}}\hskip3.8cm{\bf{(b)}}\hskip3.9cm{\bf{(c)}}\hskip3.8cm{\bf{(d)}}
{\scalebox{0.24}{\includegraphics{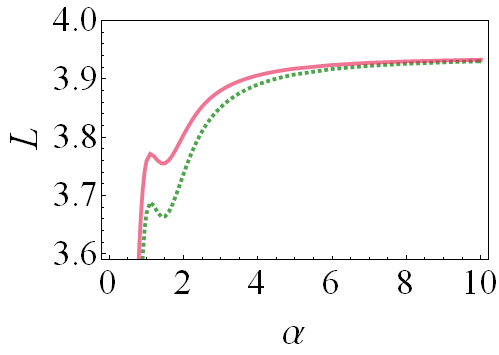}}}
{\scalebox{0.24}{\includegraphics{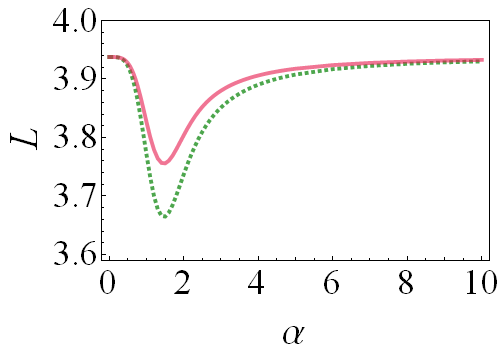}}}
{\scalebox{0.24}{\includegraphics{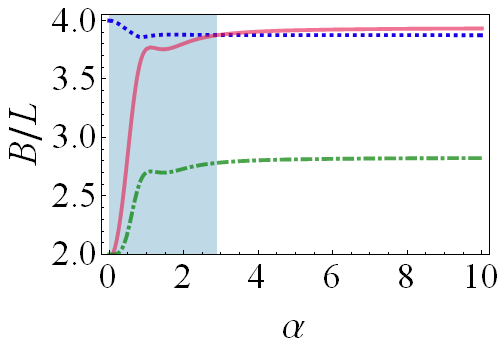}}}
{\scalebox{0.24}{\includegraphics{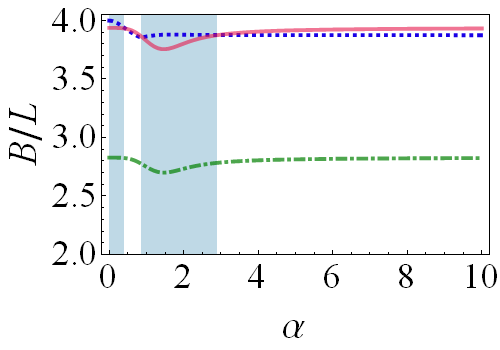}}}
\caption{(Color online)
{\bf (a)}, {\bf (b)} Leggett function (green dotted) for {\bf (a)} $\ket{\mathrm{ECS}_+}$ and 
{\bf (b)} $\ket{\mathrm{ECS}_-}$
its optimized one (red solid) as functions of $\alpha$.
{\bf (c)}, {\bf (d)} Optimized CHSH function $B$ (green dot-dashed), optimized Leggett function 
$L$ (red solid)
for {\bf (c)} $\ket{\mathrm{ECS}_+}$ and {\bf (d)} $\ket{\mathrm{ECS}_-}$ plotted, together 
with the non-local realistic bound (blue dotted), against $\alpha$.
As  the local realistic bound is $2$, the CHSH inequality is violated for any value of $\alpha$ 
within this range except at $\alpha=0$ for $\ket{\mathrm{ECS}_+}$. Leggett inequality is 
violated for $\alpha \gtrsim 2.9$.
}
\label{LeggettTest37Opt}
\end{figure*}

\section{Leggett inequality test for ECS with pseudo-spin measurements}
\label{pseudo}

The ECSs (\ref{ECS}) are known to show Bell violations
for almost any value of $\alpha$, when pseudo-spin measurements are used \cite{BellTest1}.
The pseudo-spin operators are defined as \cite{Pseudospin}
\begin{equation}
\hat s_z = (-1)^{\hat a\hat a^\dagger}, \quad
\hat s_{-}=\hat s^\dagger_{+} = \sum^{\infty}_{n=0} \ket{2n} \bra{2n+1}
\end{equation}
and satisfy the SU(2) algebra of standard spin-$1/2$ particles.
Here, $\hat s_z$ is the parity operator in Eq. (\ref{ParityOperator}).
As discussed in Ref. \cite{Pseudospin}, we need
the combined local operation and measurement observable given by
\begin{equation}
{\bf a} \cdot \hat{\bf  s} = \sin \theta (e^{i\phi}
\hat s_{-} + e^{-i\phi} \hat s_{+}) + \cos \theta \hat s_z,
\label{eq:rot}
\end{equation}
where $\mathbf{a}{=}(\sin\theta \cos\phi, \sin\theta \sin\phi,\cos\theta)$ is a measurement 
vector and
$\hat{\bf s} = (\hat s_x, \hat s_y, \hat s_z)$
is the pseudo-spin operator vector with $\hat s_{\pm}{=}\hat s_x \pm i \hat s_y$. Due to the 
bidimensional character of such local operations,
both a Bell-CHSH test and a Leggett one are possible.

The correlation function for (say) $\ket{\mathrm{ECS}_-}$ with
measurement vectors $\mathbf{a} = (\theta_A, \phi_A)$ and $\mathbf{b} = (\theta_B, \phi_B)$ is
\begin{equation}
\label{EPseudospin}
\begin{aligned}
&E_{-}(\mathbf{a,b}) {=} \bra{\mathrm{ECS}_-} (\mathbf{a \cdot s}) (\mathbf{b \cdot s}) 
\ket{\mathrm{ECS}_-}\\
&{=} -\cos\theta_A \cos\theta_B{-}K(\alpha) \sin\theta_A \sin\theta_B\cos(\phi_A{-}\phi_B)
\end{aligned}
\end{equation}
with
\begin{equation}
K(\alpha){=}\!\frac{2\alpha^2}{\sinh 2\alpha^2}
\left[ \sum^{\infty}_{n=0} \frac{\alpha^{4n}}{(2n)! \sqrt{2n+1}} \right]^2.
\end{equation}
Quantitatively, we have that $0.907 \leq K(\alpha) \leq 1$ and $K(\alpha) \rightarrow 1$ as
${\alpha\rightarrow0}$ or $\infty$. In this limit we have $E_{-}(\mathbf{a,b}) = - \mathbf{a 
\cdot b}$, which is exactly the correlation function of a PES [see Eq. (\ref{PES})].
For the case of $\ket{\mathrm{ECS}_+}$,
if the direction of one of the vectors identifying a measurement is inverted,
the correlation function becomes identical to Eq.~(\ref{EPseudospin}) with $K(\alpha)$ replaced 
by
$\tanh(2\alpha^2)K(\alpha)$.
The local average for pseudo-spin measurements can be calculated, in analogy with Eq. 
(\ref{MalusLawOnoff}), as
\begin{equation}\label{MalusLawPseudospin}
\overline{A} (\mathbf{u;a}) =
\bra{\alpha} (\mathbf{u \cdot s})^\dagger (\mathbf{a \cdot s}) (\mathbf{u \cdot 
s})\ket{\alpha}.
\end{equation}
Here, the additional pseudo-spin operator $\mathbf{u\cdot s}$
with another unit vector $\mathbf{u}$
plays the role of the rotation operator as in Eq.~(\ref{MalusLawOnoff}).
Note that $\mathbf{u\cdot s}$ is unitary
and for an arbitrary unit vector $\mathbf{a'}$,
we can get $\mathbf{u} = (\mathbf{a + a'})/\sqrt{2( 1 + \mathbf{a \cdot a'})}$ satisfying
\begin{equation}
(\mathbf{u \cdot s})^\dagger (\mathbf{a \cdot s}) (\mathbf{u \cdot s})
= \mathbf{a' \cdot s}.
\end{equation}
Clearly, the role of $\mathbf{u\cdot s}$ is to rotate the axis
of the pseudo-spin measurement from $\mathbf{a}$ to $\mathbf{a'}$.
The local operator $\mathbf{u\cdot s}$ transforms a coherent state $\ket{\pm \alpha}$
under the same assumption as in Eq.~(\ref{Transformation})
\begin{equation}
\label{TransformationEPseudospin}
\begin{aligned}
&\ket{\alpha}\rightarrow
\sin \theta \cos \phi \ket{\alpha} - (\cos \theta + i \sin \theta \sin \phi ) \ket{-\alpha},\\
&\ket{-\alpha}\rightarrow
- (\cos \theta - i \sin \theta \sin \phi ) \ket{\alpha} - \sin \theta \cos \phi \ket{-\alpha}.
\end{aligned}
\end{equation}
With Eqs. (\ref{EPseudospin}) and (\ref{MalusLawPseudospin})
we test the Leggett inequality corresponding to the ``3+6'' and ``3+7'' settings.

\subsection{3+7 setting Leggett inequality test}
\label{3+7}

In order to test Leggett inequality with an ECS, we first obtain numerically
the non-local realistic constraint $f_{\rm min}(\varphi)$ in Eq. (\ref{inequality37}).
We then compare the (non-optimized) Leggett function with such bound, as shown
in Fig. \ref{LeggettTest37} for $\ket{\mathrm{ECS}_-}$ and $\alpha=5,50$.
As expected, the bound depends on the measurement settings. Moreover, there is a dependence on 
$\alpha$ as well, although this becomes very weak as soon as $\alpha \gtrsim 5$.
Evidently, there is a range of values of $\varphi$ where the inequality is violated. The degree 
of violation is maximized at $\varphi \simeq 0.25$, regardless of $\alpha$ and for both 
$\ket{\mathrm{ECS}_-}$ and $\ket{\mathrm{ECS}_+}$. Such value agrees exactly with that 
maximizing the degree of violation when the PES in Eq. (\ref{PES}) is used. This should be 
expected as, for $\alpha \rightarrow 0$ or $\infty$,
$E_{-}(\mathbf{a,b}) \rightarrow - \mathbf{a \cdot b}$.

Retaining $\varphi = 0.25$ in our calculations,
we optimize the Leggett function by rigid-body rotations of the measurement vectors, as 
illustrated before. As it can be seen in Fig. \ref{LeggettTest37Opt} {\bf (a)}-{\bf (b)} ,
the optimization really helps Leggett function to grow larger within the range of $\alpha$ 
considered in our study, although the enhancement progressively decreases as $\alpha 
\rightarrow 0$ or $\infty$.
In the spirit of the investigations performed in Refs. \cite{LeggettCV1,LeggettCV2}, in Fig. 
\ref{LeggettTest37} {\bf (c)}-{\bf (d)} we also compare the Leggett and optimized CHSH 
functions for $\ket{\mathrm{ECS}_\pm}$ so as to elucidate the relation between the two 
inequalities.
As proven earlier in this paper, the parameter region where Leggett inequality is violated 
stays within the region where the CHSH inequality is violated too.
As also mentioned in Refs. \cite{LeggettCV1,LeggettCV2}, there is an interesting region of 
values of $\alpha$ where, while the CHSH inequality is violated, Leggett's is not.
It is worth stressing the effectiveness of the optimization procedure adopted in our analysis. 
For instance,
the minimum amplitude $\alpha$ of $\ket{\mathrm{ECS}_+}$ for the violation of Leggett 
inequality was 7.5 in Ref. \cite{LeggettCV2}, while it is lowered down to $2.9$ here, as a 
result of the optimization over rigid-body rotations. In the next Subsection, it is shown that 
for the ``3+6 settings" inequality, such threshold value is even smaller and equal to $1.8$.

\begin{figure}[t]
\centerline{ \bf{(a)} }
\centerline{\scalebox{0.30}{\includegraphics{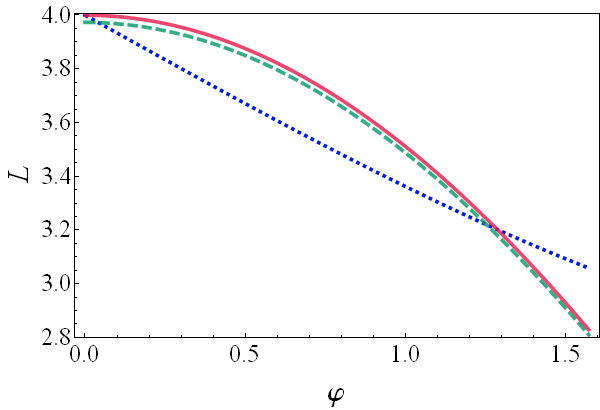}}}
\centerline{ \bf{(b)} }
\centerline{\scalebox{0.30}{\includegraphics{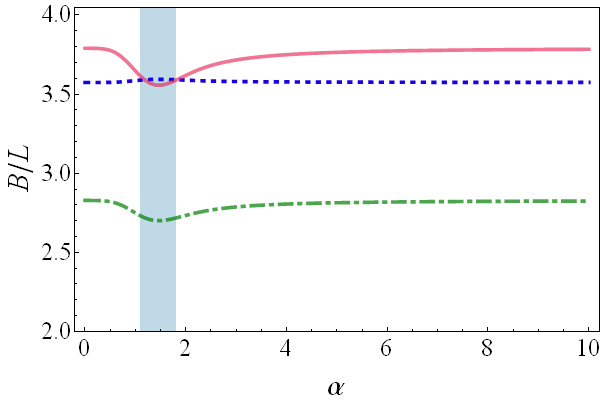}}}
\caption{(Color online)
{\bf (a)} Leggett function $L$ at $\alpha=$ 5 (green dashed) and 50 (red solid)
and the non-local realistic bound (blue dotted) as functions of $\varphi$.
The parameters used for this plot are the same as in Fig. \ref{LeggettTest37}{\bf (a)}, 
although the degree of violation of Leggett inequality is maximized at $\varphi \simeq 0.65$.
{\bf (b)} Optimized CHSH function $B$ (green dot-dashed), optimized Leggett function $L$ (red 
solid) and
the non-local realistic bound (blue dotted) for $\ket{\mathrm{ECS}_-}$.
}
\label{LeggettTest36}
\end{figure}

\subsection{3+6 setting Leggett inequality test}
\label{3+6}

We complete our study by addressing now the form of Leggett inequality given in Eq. 
(\ref{inequality36}), which needs only $6$ measurement settings at $B$ site
and needs no optimization. We follow the same procedure described above for the ``3+7 settings"
case, except that we skip the unnecessary optimization.
As it can be appreciated by examining Fig. \ref{LeggettTest36} {\bf (a)},
the value of $\varphi$ maximizing the degree of violation of the Leggett inequality  is 
$\varphi \simeq 0.65$, which agrees with
the one found for a PES [Eq. \ref{PES}] \cite{exp3}.
Here too, we retain this value and study both the CHSH and Leggett functions
for $\ket{\mathrm{ECS}_-}$ [we omit the case of $\ket{\mathrm{ECS}_+}$ for the sake of 
brevity]. Fig. \ref{LeggettTest36} {\bf (b)} shows the optimal nature of the ``3+6 settings" 
inequality: The region where Leggett inequality is satisfied is halved with the degree of 
violation being doubled as compared with the ``3+7 settings'' case.

As for the the qualitative similarity between the curves of CHSH and Leggett functions,
which is observed in common in Fig. \ref{LeggettTest37} {\bf (c)-(d)}
and \ref{LeggettTest36} {\bf (b)},
some considerations can be drawn.
First, the small dip in  found around $\alpha \simeq 1.5$ and common to all the graphs is 
solely due to the similar behavior that $K(\alpha)$ has in the two cases.
Second, one finds a similar sudden drop of the curves associated to $\ket{\mathrm{ECS}_+}$
as $\alpha \rightarrow 0$ [we only show it for the ``3+7 settings'' case in Fig. 
\ref{LeggettTest37} {\bf (c)}]. This is due to the fact that $\ket{\mathrm{ECS}_+} \rightarrow 
\ket{0,0}$ as $\alpha\rightarrow0$.
The similarity between the Leggett corresponding to the ``3+7 settings'' and ``3+6 settings'' 
is somehow to be expected, given that the two functions have been constructed using the same 
assumptions on non-local realism. On the contrary, we believe the analogies between the 
behavior of the CHSH and Leggett functions are striking in consideration of the different 
arguments at the basis of the two inequalities.

\section{Conclusion}
\label{conclusions}

through
We have performed the first step towards the construction of a formal apparatus
for the rigorous extension of Leggett's test for non-local realism to the CV scenario,
therefore going significantly beyond the efforts performed in Ref.~\cite{LeggettCV2}.
Technically, our tools have been borrowed from the considerable body of studies performed so 
far on the violation of local realism by this class of states and include on/off, parity and 
pseudo-spin measurements. The requirements for local state-definiteness at the basis of 
Leggett's arguments impose some fundamental constraints resulting in the necessity for the 
research for a new local-realistic bound, when CV states are used.
While generalizing Leggett inequality to a form suitable for ECS resources, we have 
analytically clarified the relation between the violation of the CHSH and Leggett inequality
and such relation has been
exemplified by studying the behavior of the CHSH and Leggett functions against the sizes of 
ECSs. We believe that our study contributes significantly to the understanding of the facets of 
non-local realism, in particular with respect to the generalization of Leggett's original 
formulation to entangled CV states.
Our results highlight the necessity for a more general way to construct non-local realistic 
tests
applicable to any quantum-mechanical entangled states, an intriguing task that would be 
interesting to pursue both theoretically and experimentally.

\acknowledgments

MP thanks the Center for Macroscopic Quantum Control and the Department of Physics and 
Astronomy,
Seoul National University, for kind hospitality while this work has been initiated.
This work was supported by the National Research Foundation of Korea (NRF) grant funded by
the Korea government (MEST) (No. 3348-20100018),
Center for Subwavelenth Optics (R11-2008-095-01000-0),
and the World Class University (WCU) program.
MP is supported by the UK EPSRC (EP/G004579/1).


\begin{thebibliography}{99}
\bibitem{commentosecond} A second relation also introduced by Leggett, namely
\begin{equation}\label{BasicInequality2}
-1 + \left| \overline{A} + \overline{B} \right|
\leq \overline{A B},
\end{equation}
is redundant. It is easily obtained by taking $B \rightarrow -B$ in Eq. 
(\ref{BasicInequality1})
and gives rise to the same consequences as Eq. (\ref{BasicInequality1}).
\bibitem{Derivation}
The derivation of Leggett inequality given here skips a number of technical
and non-trivial steps whose presentation is immaterial for the tasks and purposes of our work.
The reader interested in such details could consult the original paper by Leggett 
\cite{Leggett} as well as
Refs. \cite{exp1,exp2,exp3}.
\end{thebibliography}
\end{document}